\documentclass[prl,aps,twocolumn, floatfix, superscriptaddress,showpacs]{revtex4-1}
\usepackage{amsmath,bm,graphicx}
\usepackage{amssymb}
\usepackage{amsfonts}
\usepackage{color}

\newcommand{\pd}[2]{\frac{\partial #1}{\partial #2}}
\newcommand{\dd}[2]{\frac{d #1}{d #2}}

\begin{document}
\title{Some remarks on the inverse Smoluchowski problem for cluster-cluster aggregation}
\author{Colm Connaughton}
\email{connaughtonc@gmail.com}
\affiliation{Mathematics Institute, University of Warwick, Coventry CV4 7AL, UK}
\affiliation{Centre for Complexity Science, University of Warwick, Coventry CV4 7AL, UK}

\author{Peter P. Jones}
\affiliation{Centre for Complexity Science, University of Warwick, Coventry CV4 7AL, UK}
\email{P.P.Jones@warwick.ac.uk}
\date{\today} 
 
\begin{abstract}
It is proposed to revisit the inverse problem associated with Smoluchowski's coagulation equation. The objective 
is to reconstruct the functional form of the collision kernel from observations of the time evolution of 
the cluster size distribution. A regularised least squares method originally proposed by Wright and 
Ramkrishna (1992) based on the assumption of self-similarity is implemented and tested on numerical data 
generated for a range of different collision kernels. This method expands the collision kernel as a sum of 
orthogonal polynomials and works best when the kernel can be expressed exactly in terms of these polynomials. It 
is shown that plotting an ``L-curve'' can provide an a-priori understanding of the optimal value of the 
regularisation parameter and the reliability of the inversion procedure. For kernels which are not exactly 
expressible in terms of the orthogonal polynomials it is found empirically that the performance of the method 
can be enhanced by choosing a more complex regularisation function.
\end{abstract}

\maketitle

\section{Introduction}

The effects of clouds and precipitation is one of the largest sources of uncertainty in our
current attempts to simulate the Earth's climate \cite{STE2005}. The reason for this is that most phenomena 
associated with clouds happen on scales below those which are explicitly resolved by the current climate
models. Their feedback on resolved scales must therefore be parameterised. Such parameterisations require a strong
understanding of the underlying physical processes taking place. One such process which has attracted
considerable attention in recent years is the time evolution of the droplet size distribution in 
clouds and its connection with precipitation. Water droplets in clouds are seeded by cloud condensation
nucleii such as aerosol particles. They initially grow by condensation and subsequently by coagulation as
droplets collide with each other in the cloud to produce larger droplets. The detailed micro-physics of the 
coagulation process is difficult to understand theoretically since turbulence within the cloud is
believed to play a key role in determining the collision rate of droplets \cite{BMSS2010}. A 
complete theoretical description is hampered by the difficulty in describing the statistical interplay
between particles and turbulence analytically. On the other hand the quality of the data available for
the study of this problem is improving rapidly due to recent advances in 
observational techniques \cite{SLWF2006} and direct numerical simulation of droplets in turbulent flows 
\cite{RC2000,WARG2008}. 

In the context of the droplet coagulation problem, much research has focused on using our improved 
observational understanding of the behavious of droplets in turbulent flows to calculate more accurately
the functional form of the droplet coagulation rate, $K(m_1,m_2)$, as a function of droplet masses, or
 equivalently (if droplets are assumed spherical), droplet size. In this note, we argue that the
improved availability of data suggests that we should also revisit the corresponding inverse problem which
can be stated as follows: given observations or measurements of the time evolution of the droplet size 
distribution, how much information can be extracted about the functional form of the coagulation rate?
This problem was studied in the past by Wright and Ramkrishna \cite{WR1992} in the context of 
chemical mixing and has been recently revisited in the context of droplets in turbulence by
Onishi et al. \cite{OMTKK2011}. Although there are many difficulties associated with such inverse problems,
as discussed below, there are potential rewards. The approach is data driven and could provide a guide
to modeling in situations where the underlying microphysics remains incompletely understood. The inverse
approach could also begin to quantify the extent to which available data and measurements can distinguish
between different models.

First a word about the usual forward problem. If the collision kernel, $K(m_1,m_2)$, is known, the evolution of 
the cluster size distribution, $N_m(t)$ is
described by the Smoluchowski coagulation equation \cite{SMO1917}:
\begin{eqnarray}
\nonumber \partial_t N_m(t) &=& \int_0^{m}dm_1\, K(m_1,m-m_1)\, N_{m_1}(t)\, N_{m-m_1}(t)\\
\label{eq-Smoluchowski}  &-&2 N_m(t)  \int_0^{\infty}dm_1\, K(m,m_1)\, N_{m_1}(t).
\end{eqnarray}
It is applicable when spatial correlations between particles are sufficiently weak that collisions between
particles can be considered statistically independent. A huge amount is known theoretically about the 
solutions of the Smoluchowski equation. See \cite{LEY2003} for a modern review.

\section{Homogeneous collision kernels and self-similarity}
\label{eq-selfSimilarity}

\begin{figure}
\includegraphics[width=6.5cm]{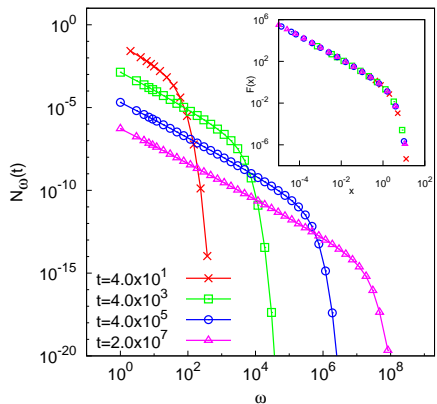}
\caption{\label{fig-scaling} Snapshots of the self-similar evolution of the cluster size distribution, $N_m(t)$, obtained from a numerical simulation of Eq.~(\ref{eq-Smoluchowski})  with the collision kernel $K(m_1, m_2) = (m_1,m_2)^\frac{1}{4}$. The inset shows the scaling function obtained when the data is collapsed according to Eq.~(\ref{eq-SS}) with the typical scale, $s(t)$, obtained as in Eq.~(\ref{eq-sDefn}) with $n=2$. }
\end{figure}

In some applications (see \cite{ERN1986} for some discussions), the collision kernel is a homogeneous function of its 
arguments whose degree we shall denote by $\gamma$:
\begin{displaymath}
K(a m_1, a m_2)= a^{\gamma}\,K(m_1,m_2).
\end{displaymath}
For such kernels, the evolution of the cluster size distribution is often self-similar. That is to say we can
write:
\begin{equation}
\label{eq-SS}
N_m(t) \sim s(t)^{-2}\, f\left(z\right) \hspace{1.0cm} z=\frac{m}{s(t)}
\end{equation}
where $s(t)$ is the typical cluster size. This can be defined intrinsically as a ratio of moments of the size 
distribution:
\begin{equation}
\label{eq-sDefn}
s(t) = \frac{M_{n+1}(t)}{M_{n}(t)}\hspace{1.0cm}M_p(t) = \int_0^\infty m^p\,N_m(t)\,d m.
\end{equation} 
The scaling function, $f(z)$, determines the shape of the cluster size distribution. An example of this
self-similar evolution obtained from numerical simulation of Eq.~(\ref{eq-Smoluchowski}) with 
$K(m_1,m_2) = (m_1\,m_2)^{1/4}$ is shown in Fig.~\ref{fig-scaling}. The inset shows the scaling function, 
$f(z)$, obtained by collapsing the data according to Eq.~(\ref{eq-SS}) with the typical scale, $s(t)$, obtained
as in Eq.~(\ref{eq-sDefn}) with $n=2$. All numerical simulations of Eq.~(\ref{eq-Smoluchowski}) reported
in this paper were done using the pairwise binning method described in \cite{LEE2000}.

A homogeneous collision kernel and self-similar time evolution are not plausible assumptions for the droplet 
coagulation problem in clouds which has motivated this study \cite{FFS2002}. If one restricts attention
to the gravitational coagulation-dominated regime, one could perhaps argue that the kernel is 
approximately homogeneous. It turns out, however, that even with the resulting differential sedimentation
kernel which is homogeneous of degree $\gamma=4/3$,  the Smoluchowski equation does not produce a self-similar 
evolution but rather undergoes instantaneous gelation \cite{VDO1987,BCSZ2010-arxiv}. Our focus on
self-similar problems simply stems from the fact that they provide the simplest
class of coagulation problems on which we can begin to study the inverse problem described above. We
further restrict ourselves to kernels having $\gamma<1$ in order to avoid dealing with the gelation
transition.

\section{The inverse Smoluchowski problem}
\label{sec-inverseProblem}

Let us introduce the cumulative cluster mass distribution:
\begin{equation}
\mathcal{F}_m(t) = \int_0^m m_1\,N_{m_1}(t) d m_1.
\end{equation}
The original cluster size distribution, $N_m(t)$ is recovered from $\mathcal{F}_m(t)$ by differentiation:
\begin{equation}
N_m(t) = \frac{1}{m}\,\pd{\mathcal{F}_m(t)}{m}.
\end{equation}
In terms of $\mathcal{F}_m(t)$, Eq.~(\ref{eq-Smoluchowski}) can be written in the compact form:
\begin{equation}
\partial_t \mathcal{F}_m(t) = - \int_0^{m} d \mathcal{F}_{m_1}(t) \int_{m-m_1}^\infty \frac{d \mathcal{F}_{m_2}(t)}{m_2}\, K(m_1,m_2) 
\end{equation}
If we assume scaling as in Eq.~(\ref{eq-SS}) then $\mathcal{F}_m(t) = F(z)$ and the scaling function of
the cumulative mass distribution, $F(z)$, satisfies the following scaling equation:
\begin{equation}
\label{eq-scalingEquation}
z \dd{F}{z} = - \int_0^{z} d F(z_1) \int_{z-z_1}^\infty \frac{d F(z_2)}{z_2}\, K(z_1,z_2).
\end{equation}
This is a complicated nonlinear integro-differential equation if we are required to determine $F(z)$ given 
$K(z_1, z_2)$) (the forward problem). It is, however, a linear equation if we are required to determine 
$K(z_1, z_2)$ given $F(z)$ (the inverse problem). This inverse problem is, however, ill-posed. This is
most easily seen by considering what happens if we discretise Eq.~(\ref{eq-scalingEquation}) on $N$ $z$-points.
We obtain a linear system,
\begin{displaymath}
\mathbf{g} = S\, \mathbf{k},
\end{displaymath}
consisting of $N$ equations for the $N^2$ values of $K(z_1, z_2)$ on the discretisation points. Actually we
can reduce the number of unknowns almost by a factor of 2 since $K(z_1, z_2) = K(z_2,z_1)$ but the 
conclusion remains unchanged. This system is enormously under-determined, which implies that one can find many 
solutions but they are typically over-fitting the measurements of $F(z)$.

One way of dealing with under-determined systems is via Tikhonov Regularisation (Ridge regression). The
idea is to solve a minimization problem with a regularisation term:
\begin{equation}
\label{eq-minimisation}
\mathbf{k}_\mathrm{est} = \arg\min_{\mathbf{k}}\, \left\{ \| S\,\mathbf{k} - \mathbf{g}\|^2 + \lambda \|\mathbf{k}\|^2 \right\}.
\end{equation}
Noise-dominated solutions have to compete against the regularization term $\lambda \|\mathbf{k}\|^2$
in the minimisation. This approach was pioneered by Wright and Ramkrishna \cite{WR1992}. It has the advantage
that, in principle, one does not need to know the functional form of the collision kernel a-priori.
The trick is to choose the ``best'' value of the regularization parameter, $\lambda$. Before discussing the
selection of the value of $\lambda$, we discuss briefly how to set up the minimization problem.

Following \cite{WR1992}, we assume that $K(m_1, m_2)$ can be expanded in terms of Laguerre polynomials, $L_i(x)$,
 up to order $p$ in the $m_1$ and $m_2$ directions:
\begin{equation}
\label{eq-kernelExpansion}
K(m_1,m_2) = \sum_{n=1}^{p^2} a_n \, L_n(m_1, m_2)
\end{equation}
where
\begin{displaymath}
L_n(x,y) = L_i(x)\,L_j(y)
\end{displaymath}
where $n=(i-1)\,p + j$ is a compound index which rolls the indices, $i$ and $j$, in the $m_1$ and $m_2$ 
directions into one. With Eq.~(\ref{eq-kernelExpansion}), the minimisation problem Eq.~(\ref{eq-minimisation})
is transformed into a different minimisation which aims to determine the coefficients, $a_n$ in 
Eq.~(\ref{eq-kernelExpansion}). This has a number of advantages. It  ensures that the minimisation problem 
automatically finds kernels which are fairly smooth functions of $m_1$ and $m_2$. It also enormously 
reduces the size of the problem since the number of polynomials, $p$, required in each direction is
typically rather small. Furthermore, this choice allows the entries of the matrix $S$ in 
Eq.~(\ref{eq-minimisation}) to be calculated semi-analytically as described in \cite{WR1992}. It has the 
disadvantage, however, of constraining the solution to those kernels which 
can be expressed in the form of Eq.~(\ref{eq-kernelExpansion}) which effectively requires
a-priori knowledge of the class of kernels which we are seeking, as was done in \cite{OMTKK2011}.

As mentioned above, the tricky part of this procedure is to determine the most apprpriate value of $\lambda$
to use in Eq.~(\ref{eq-minimisation}). If $\lambda$ is too small, the result will over-fit the observations. If
$\lambda$ is too large then $\mathbf{k}_\mathrm{est}$ will be pushed to zero by the weight of the second term in
Eq.~(\ref{eq-minimisation}) and will retain little information about the Smoluchowski dynamics encoded in the
matrix $S$. A rational approach to determining $\lambda$ is provided by plotting an ``L-curve'' (for a
 clear review see \cite{HAN2001}). 
This is a plot of the norm of the solution, $\|\mathbf{k}\|$, as a function of the norm of the
residual, $\| A\,\mathbf{k} - \mathbf{b}\|$, for a range of values of $\lambda$. If there is a 
clear transition in the minimisation problem from a regime dominated by overfitting to a regime dominated
by the regularisation then the L-curve will have the distinctive L-shape and the best values of 
$\lambda$ are those in the ``elbow'' of the curve.

\section{Results}

\begin{figure*}
\begin{tabular}{cc}
\includegraphics[width=6.5cm]{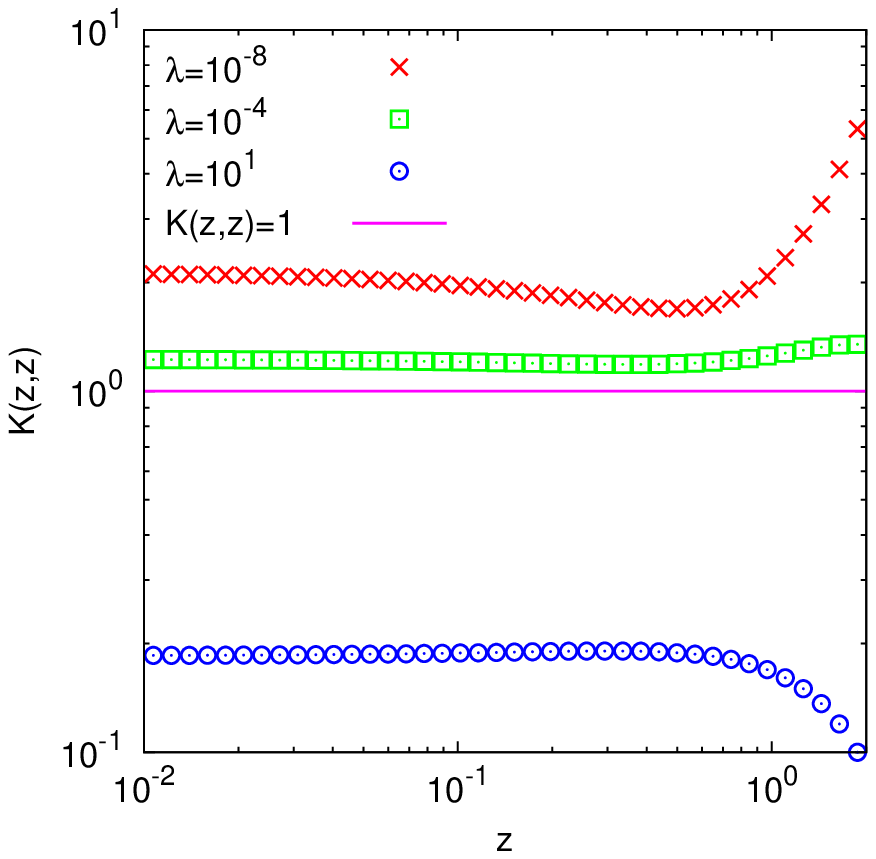}
&
\includegraphics[width=6.5cm]{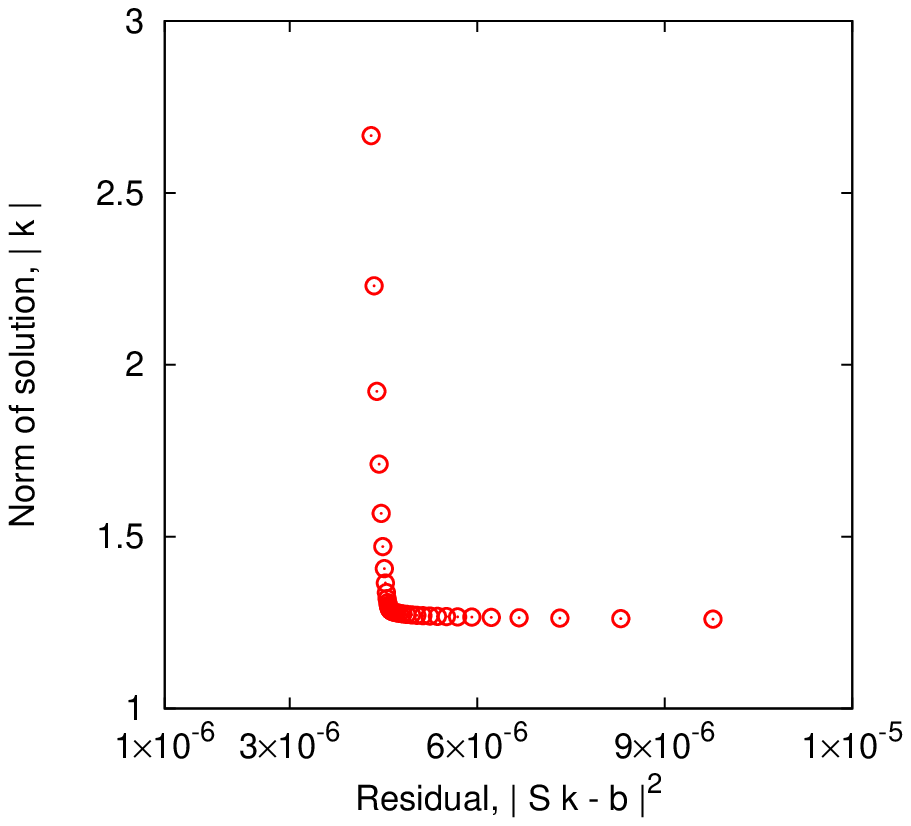}
\end{tabular}
\caption{\label{fig-resultsConstantKernel} Results for the constant kernel, $K(m_1,m_2=1$. The right panel
shows the L-curve obtained by performing the minimisation Eq.~(\ref{eq-minimisation}) over a range of
values of $\lambda$. The left panel shows the diagonal, $K(z,z)$, of the reconstructed kernels compared to
the theoretical curve for values of $\lambda$ in the upper left, lower right and the ``elbow'' of the L-curve.}
\end{figure*}

\begin{figure*}
\begin{tabular}{cc}
\includegraphics[width=6.5cm]{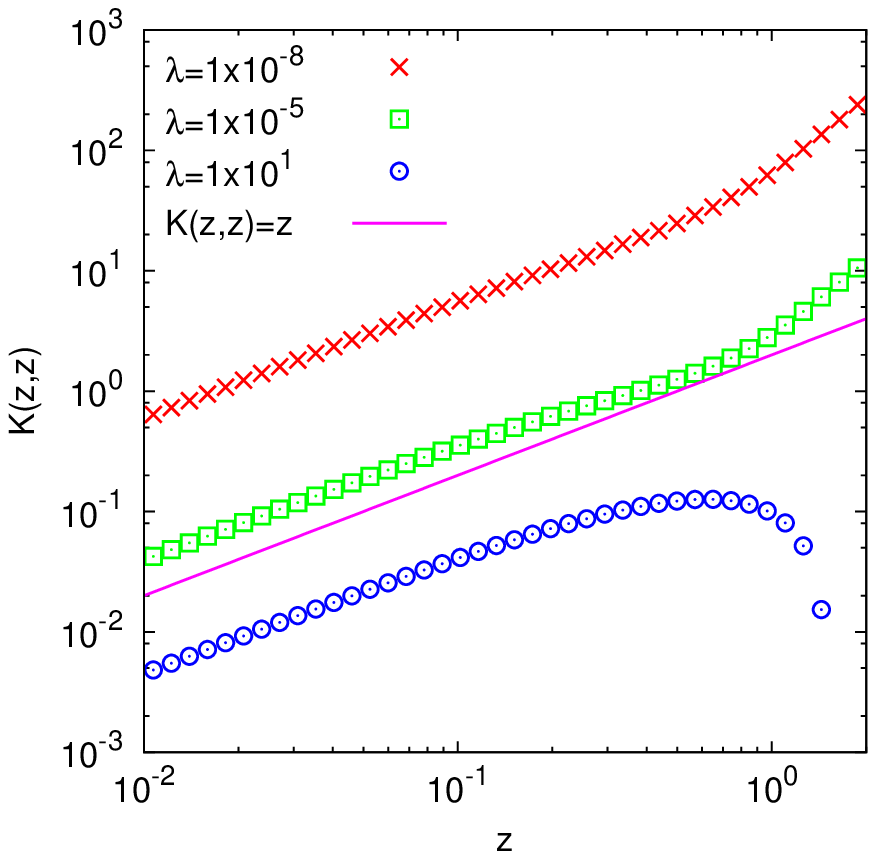}
&
\includegraphics[width=6.5cm]{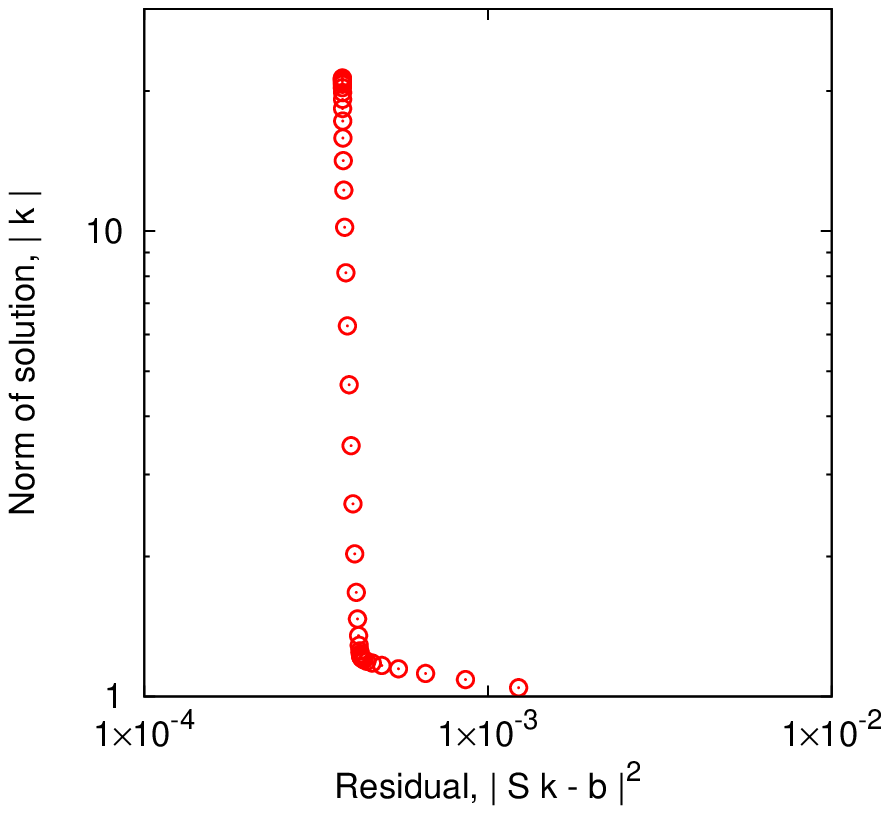}
\end{tabular}
\caption{\label{fig-resultsSumKernel} Results for the sum kernel, $K(m_1,m_2=\frac{1}{2}(m_1+m_2)$. The right panel
shows the L-curve obtained by performing the minimisation Eq.~(\ref{eq-minimisation}) over a range of
values of $\lambda$. The left panel shows the diagonal, $K(z,z)$, of the reconstructed kernels compared to
the theoretical curve for values of $\lambda$ in the upper left, lower right and the ``elbow'' of the L-curve. }
\end{figure*}

\begin{figure*}
\begin{tabular}{cc}
\includegraphics[width=6.5cm]{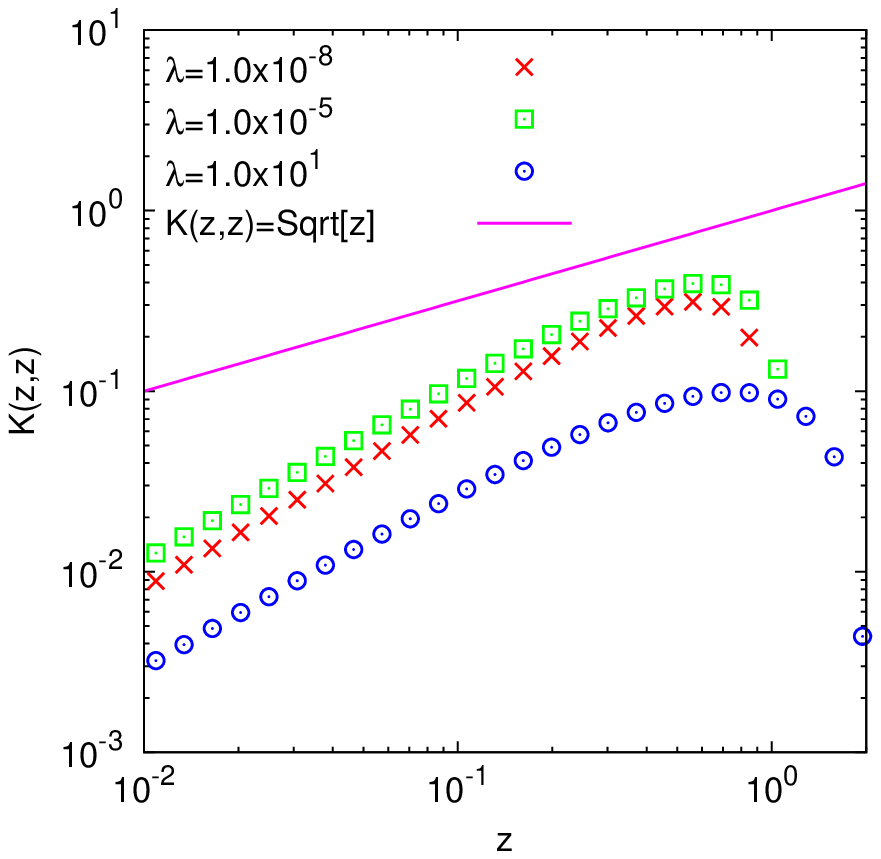}
&
\includegraphics[width=6.5cm]{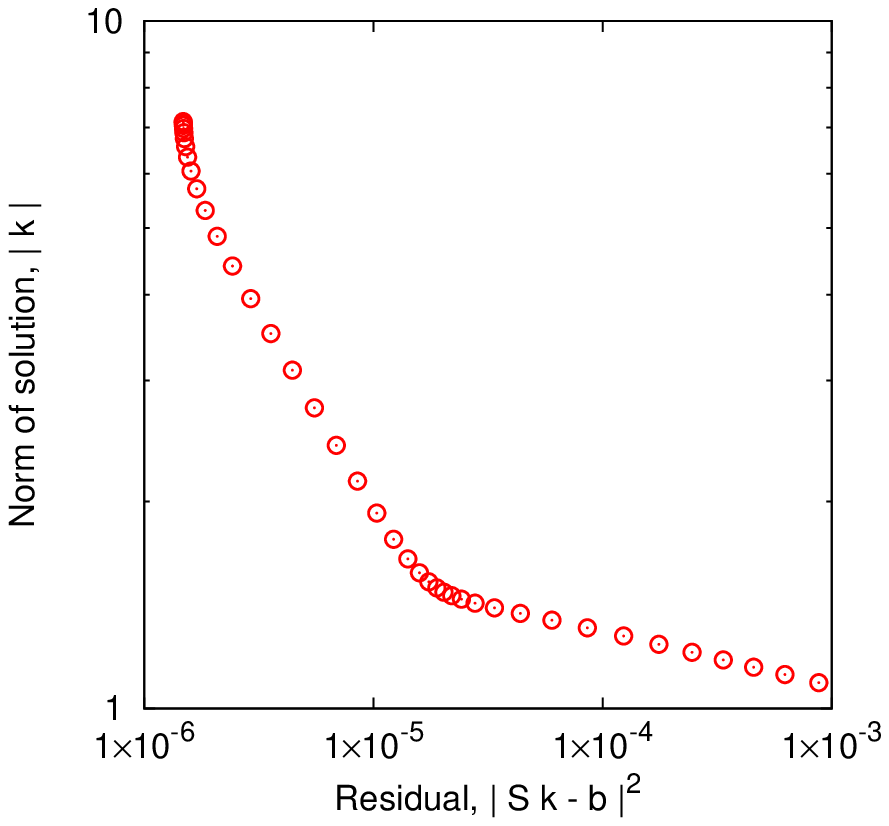}
\end{tabular}
\caption{\label{fig-resultsgenSumKernel0.5} Results for the generalised sum kernel, $K(m_1,m_2=\frac{1}{2}(\sqrt{m_1}+\sqrt{m_2})$. The right panel
shows the L-curve obtained by performing the minimisation Eq.~(\ref{eq-minimisation}) over a range of
values of $\lambda$. The left panel shows the diagonal, $K(z,z)$, of the reconstructed kernels compared to
the theoretical curve for values of $\lambda$ in the upper left, lower right and the ``elbow'' of the L-curve.  }
\end{figure*}

The regularised least squares method described above was applied to several sets of data obtained by the
numerical integration of Eq.~(\ref{eq-Smoluchowski}) with monodisperse initial data with different model 
kernels. We used 40 discretisation points in the interval $z\in \left[0.01, 2.0\right]$ and chose $p=3$
Laguerre polynomials in each direction. In each case, the data
collapse and extraction of the scaling function was done using the method described in \cite{BS2001} and
then fitted to a function of the form $C\,z^{\alpha-1}\,e^{-\beta\,z}$ as suggested in \cite{WR1992}. These
fitted scaling functions were then used as the ``observations'', $F(z)$, in the discretisation of 
Eq.~(\ref{eq-scalingEquation}).

Figs.~\ref{fig-resultsConstantKernel} and \ref{fig-resultsSumKernel} show some results of the regularised
least squares for the constant kernel, $K(m_1, m_2) =1$, and the sum kernel, $K(m_1,m_2) = \frac{1}{2}(m_1+m_2)$,
respectively. These are good test cases to begin with since the solution of Eq.~(\ref{eq-Smoluchowski})
with monodisperse initial data is known explicitly for each of these kernels allowing the numerical
aspects of the calculation to be validated. The right panels of the figures show the L-curves obtained
by performing the minimisation (\ref{eq-minimisation}) over a range of values of $\lambda$. In both
cases, a clear elbow is visible in the resulting curve indicating an appropriate range of values 
for $\lambda$. The left panel shows, for ease of visualisation, the diagonal, $K(m,m)$, of the kernels
obtained by regularised least squares as a function of $m$ for different values of $\lambda$. The $\lambda$
curves shown correspond to values of $\lambda$ which are too big, too small and ``just right'' (meaning
a value in the elbow of the L-curve). It is clear that the method does a reasonable job of extracting the
basic shape of the kernel given the scaling function, $F(z)$. Surprisingly, we found that it was not necessary
to explicitly enforce the positivity of $K(m_1, m_2)$ by constrainting the minimisation as was done in
\cite{WR1992}. The method seemed to do equally well, and in some cases, better without these constraints.
For the sum kernel, we did enforce the constraint that the kernel should vanish for particles of
zero mass which seems physically reasonable -  the results were less convincing without this constraint.

Fig.~\ref{fig-resultsgenSumKernel0.5} shows the corresponding results for the generalised sum kernel
$K(m_1,m_2) = \frac{1}{2}(\sqrt{m_1}+\sqrt{m_2})$. The results are clearly less convincing. This is 
probably related to the fact that this kernel cannot be expressed in terms of the Laguerre polynomials
chosen to represent the kernel in Eq.~(\ref{eq-kernelExpansion}). It is worth pointing out that the 
corresponding L-curve does not have a sharp elbow indicating that the method does not find a clear
``best'' value of $\lambda$. This seems to illustrate a point in favour of such methods - the fact that
the L-curve does not have a sharp elbow provides an a-priori indication that the results of the 
minimisation should be treated with caution.

Empirical investigation suggest that the results for the generalised sum kernel can be improved by tinkering
with the regularisation procedure. If, instead of, Eq.~(\ref{eq-minimisation}), we perform the minimisation
\begin{align}
 \min_a & \lVert S \mathbf{k} - \mathbf{g} \rVert_2^2 + \lambda\,w(\mathbf{k})
\end{align}
where
\begin{align}
   w(\mathbf{k}) & = \sum_i \log(\lvert k_i \rvert + 1) = \log \left[ \prod_i (\lvert k_i \rvert + 1)_i \right]
\end{align}
we found that the results were much better. This form for $w$ was found by experimentation. At present, this 
statement is at the level of empirical observation and requires further investigations.

\section{Discussion and outlook}

We conclude, as several previous authors have done, that the inverse Smoluchowski problem is technically 
feasible and could potentially be developed into a useful tool for the study of droplet size distributions
in clouds and other applications where the underlying microphysics is still incompletely understood. The
results presented here indicate that the L-curve provides a useful complementary tool to the methods
developed by Wright and Ramkrishna \cite{WR1992} to allow the regularisation parameter to be selected
a-priori in situations where the collision kernel is not known from the outset. It is worth mentioning that
the results presented in Figs.~(\ref{fig-resultsConstantKernel})-(\ref{fig-resultsgenSumKernel0.5}) do
not give a good indication to the casual reader of the degree of numerical sensitivity required in tackling
these problems. It became clear to us during these investigations that the ill-posedness of 
the inverse problem represented by Eq.~(\ref{eq-scalingEquation}) requires that great care be taken in
interpreting the outputs of these methods. It is also clear that further research is required, even in the
simple case of self-similar time evolution, in order to make the method more robust. Our results suggest that
we should consider more general functional forms for the collision kernel than Eq.~(\ref{eq-kernelExpansion})
in order to improve this robustness. This will probably not pose much difficulty since the analytic
simplification obtained by the use of Laguerre polynomials in \cite{WR1992} is probably less important
nowadays owing to the increased computational power which can be brought to bear on the computation
of matrix elements by quadrature when closed analytic forms are not available.

In the long run, however, it is clear that it is necessary to free ourselves of the assumptions of 
homogeneity of the kernel and self-similarity of the cluster size distribution. Some strong progress in
this direction has already been made recently by Onishi et al. \cite{OMTKK2011} who have been able to
infer the relative importance of the turbulent and gravitational coagulation as a function of 
Reynolds number from direct numerical simulations of droplet-laden turbulence using inverse methods. Our 
approach differs slightly from this work in the sense that we would like, as far as possible, to 
learn the functional shape of the kernel from the data by allowing considerable freedom in the
class of possible kernel functions. This approach would be more appropriate in situations when the
underlying micro-physics is unknown or controversial.

\end{document}